# Do Redshifted Cosmological Photons Really Violate the Principle of Energy Conservation?

Alasdair Macleod
University of the Highlands and Islands
Lews Castle College
Stornoway
Isle of Lewis
UK
HS2 0BW
Email: Alasdair.Macleod@lews.uhi.ac.uk

**ABSTRACT:** Although the Universe is far from understood, we are fairly confident about some key features: Special Relativity (SR) describes the kinematics of inertial frames; General Relativity (GR) explains gravitation; the Universe had a beginning in time and has been expanding since. Nevertheless it is quite difficult to see the 'big picture', although the idea of applying GR to the entire Universe has been very successful with a model emerging that is consistent with observation. One unpleasant feature of the model is that cosmological photons appear not to conserve energy, and the only explanation forthcoming is the claim that GR is exempt from the principle of energy conservation. It is demonstrated here that cosmological observations may legitimately be projected onto flat spacetime and can then be treated Special Relativistically, whereupon energy conservation is restored. This is not to say that the concordance General Relativistic cosmological model is incorrect, just that in observational terms there is no energy conservation anomaly.

**INTRODUCTION**

In the standard cosmological model, the Universe expands outwards from an initial 'big bang' but is subsequently restrained by its own gravitation, a process conveniently visualised through the analogy of an inflating balloon. Light from distant objects will therefore propagate along curved geodesic paths. Unfortunately this inference is contradicted by observation - recent data from the WMAP probe has determined to an accuracy of ±2% that spacetime is spatially flat [1]. The inconsistency is referred to as the flatness problem and is conveniently resolved by postulating an inflationary phase of expansion where the radius of the early Universe grew dramatically in a very short time. Now, although the entire Universe is still globally curved, the observable Universe represents only a tiny region on the 'surface', a region whose topology will consequently tend towards flatness [2]. Inflation is widely accepted as the solution to the flatness problem. However, inflation has Special Relativistic consequences that are often glossed over, but nevertheless merit consideration. Note firstly that flat spacetime is described by the Minkowski metric tensor of Special Relativity (SR) and the relationship between events in this frame will always conserve energy and momentum and indicate a constant value for the speed of light. Because spacetime is globally flat, it is inevitable that our observation platform as we look out on the Universe is an inertial frame – how then are cosmological events[1] to be mapped onto this frame?

This is an important question. An earlier paper [3] considered this issue and concluded that cosmological events must map onto the observer Lorentz frame in a way that will give a consistent value for the speed of light and thereby ensure energy conservation. In other words, there is no mechanism in SR for treating cosmological events differently to local events and still respect the well-established properties of inertial frames. The result was justified by the procedure of matching cosmological sources with proper sources moving in such a way as to continuously track the cosmological sources. We can compare the exchange of a photon between two cosmological entities and between two proper velocity entities: Consider one specific transaction where a photon is emitted by a proper source with event coordinates ($r_0$, $t_0$) and absorbed by another with event coordinates ($r_1$, $t_1$). If the shadow cosmological source emits a photon, also with event coordinates ($r_0$, $t_0$), will the photon be received by the cosmological entity located at position $r_1$ at time $t_1$? If the answer is **no** then a change in the measured speed of light for cosmological events is implied. Of course, we are operating in a geometrical spacetime framework, and a variation in the speed of light in turn implies curvature – cosmological photons will act as if spacetime is curved. However, spacetime is known to be flat out to huge cosmological distances [1]. We have a contradiction. The answer to the question must therefore be **yes**: Local and cosmological photons must propagate in spacetime in the same way and the observed events, whether local or cosmological, are consequently indistinguishable. This assessment is quite at odds with Davis and Lineweaver [4] who write:

> "In the ΛCDM concordance model all objects with redshift greater than $z \sim 1.46$ are receding faster than the speed of light. This does not contradict SR because the motion is not in the observer's inertial frame. No observer ever overtakes a light beam and all observers measure light locally to be travelling at $c$."

---

[1] There are events associated with objects subject to the cosmological expansion relative to the observer.

It is not stated how they are able to claim the observation frame is not inertial given that spacetime is globally flat and their claim must really be supported by a reasoned argument - it is insufficient merely to state that cosmological events are governed by GR and therefore not subject to SR. Without a valid counter argument, we may conclude that cosmological entities can be interpreted as having a proper (or peculiar) velocity and a redshift arising from the associated Doppler effect. Consequently, all cosmological data should be consistent with a Special Relativistic interpretation (without necessarily compromising the underlying General Relativistic model that may control the dynamics).

One must be aware that this entire issue really only arises as a consequence of inflation, for without inflation there is no question of cosmological events being observable from an inertial frame in any sensible cosmology[2].

**SPECIAL RELATIVISTIC COSMOLOGY**

The correct SR treatment[3] of cosmology is completely equivalent to the equations for the empty Milne Universe [5], which are found to deviate from observation in a small but significant manner around $z \sim 1$[4]. In light of the discussion in the previous section, the deviation is surprising, but there are many possible causes. These include evolutionary effects associated with 'standard candles', nett gravitation and a consequent deviation from flatness, the influence of transient non-geometric forces (dark energy, for example), boundary conditions [3], and specific SR effects. Much more data is required before an explicit analysis can commence; any attempt to do so now would be merely speculative. In any event, there is sufficient flexibility in the SR interpretation to deal with the apparent deviation from observation. This is discussed further in *Appendix 1*.

From the Milne SR treatment of the problem, a distinct form of the Hubble law naturally emerges [3]:

$$v_H = \frac{r_d}{T_e}. \qquad (1)$$

$r_d$ is the proper distance when the photon is emitted and $T_e$ is the proper time, both in the frame of the observer. It is easily shown that velocities defined in this way behave as proper velocities. This will naturally follow from the internal consistency of SR, but is nevertheless proven in *Appendix 2*. With this definition of a proper recession velocity, there is no conceivable cosmological experiment that will show a contradiction or paradox associated with cosmological events. For example, consider a distant cosmological source detected through highly redshifted photons. If a filter happens to be placed at any in-between point (also subject to the cosmological expansion) then the source photon frequencies blocked in the frame of the filter and the source photon frequencies the observer calculates should be blocked by the filter are consistent. There is no situation where the observer will presume the filter should act but will actually not operate in the expected way in the rest frame of the filter, thereby allowing photons to mysteriously pass through. The absorbing gas and ion clouds in the Universe are a practical demonstration of this effect. Of course, the hypothetical filter could also be a grating reliant on quantum or wave effects..

---

[2] General relativity (GR) encompasses SR and a correspondence exists where the GR curved spacetime metric tends towards the Minkowski tensor of SR in the limiting case of gravity weakening and spacetime becoming flat. Consequently, spacetime is asymptotically special relativistic. The correspondence extends to simple cosmological big bang models in the absence of inflation: If we consider a cosmological entity a fixed proper distance $d$ away from the observer, then over absolute time $T$ as the radius of the Universe increases and the global-curvature decreases, the observed recessional velocity decreases at the same time ($\approx d/T$). Thus as the metric tends towards SR over any cosmological scale, the recessional velocity (which has a GR origin) tends to disappear as might be expected. The cosmological expansion is therefore seen to be a purely general relativistic effect that disappears as the metric becomes Minkowskian.
[3] Davis and Lineweaver [3] present an incorrect description of SR cosmology.
[4] The discrepancy arises with type Ia supernovae, the WMAP data and possibly lensing data.

The requirement of consistency for a single observer is easily satisfied but the stronger condition that there should be no way of distinguishing an expansion velocity from a proper velocity[5] through any procedure using colluding observers may also hold. One can envisage a number of coupled observers (tethered in some way, perhaps) surrounding and correlating the data from a cosmological object, but we run into difficulties in deciding if observers can 'surround' a cosmological entity without themselves being subject to the expansion. However, there is one facet of the cosmological expansion that apparently makes it quite distinct from a proper velocity and we can perhaps exploit this to distinguish cosmological events: This is the apparent violation of energy conservation as photons subject to the cosmological expansion are continuously redshifted. The usual example quoted is that of the huge flux of CMBR photons, but the situation is the same for all cosmological photons. Hubble stated [6]:

> "…redshifts, by increasing wavelengths, must reduce the energy in the quanta. Any plausible interpretation of redshifts must account for the loss of energy."

It is important to recognise that, in contrast, the normal Doppler effect conserves energy. We can verify this by considering a spherical surface of static observers centred on a stationary isotropic radiating source with integrated luminosity $L$. If now the source is boosted to a velocity $v$ in an arbitrary direction relative to the shell of observers, then observers in the general direction $\hat{v}$ will measure the photons as blueshifted and observers in the general direction $-\hat{v}$ will see the photons redshifted with the total integrated luminosity still $L$. This does not imply some sort of collusion between the blue- and red-shifted photons to ensure energy conservation – in fact all events individually conserve energy. This is easily demonstrated by a 'before and after' SR kinematic analysis [7]. The energy difference between the emitted energy and the absorbed energy in excess to the standard energy difference associated with the non-zero relative velocity is fully accounted for by momentum conservation (with an apparent recoil at the source as viewed from the observer).

The Doppler explanation does not work with the conventional definition of the cosmological redshift since the sources are not considered to move through space. Photons are assumed to propagate through expanding space, losing energy as they progress. It is therefore a dynamic effect. The Doppler effect is kinematic, a function only of the emitter and absorber 4-momenta. In SR, it is unnecessary to describe photons as travelling through the medium; they are completely characterised by the emission and reception events[6]. However, if cosmological events are mapped onto flat Minkowski spacetime in the consistent way we have claimed above, the Doppler explanation then becomes applicable and the principle of energy conservation is restored. It then follows that apparent violation of energy conservation cannot be used to distinguish between local and cosmological events, and that there is no way to distinguish between local and cosmological events in SR.

We can show the CMBR photons and sources have a 'well-behaved' SR analogue by a simple argument. Consider again a stationary isotropic source surrounded by a spherical shell of observers, but this time the shell is moving at proper velocity $v$ away from the centre. All received photons will now appear redshifted and one might naively assume there to be an apparent total energy loss[7]. However this is an SR situation and there is in fact no energy loss – to see this, we merely have to make a clear distinction between the energy difference associated with a relative velocity (a rotation of the 4-momentum vector) and genuine energy loss. In the test situation we have set up, there are no forces or accelerations, so the problem can be treated in Special Relativistically. Let us assume that we can slow down the emission rate to such a level that only one photon is emitted at a time in a random direction. The observer receiving a particular photon will expect and detect an energy difference because of the emitter and absorber inertial frames are moving with a relative velocity $v$. The observer will also include in the analysis the recoil at the source. Energy is conserved and, with respect to any particular observer, the source will appear to vibrate over time as it recoils in random directions, in a sense exhibiting an increased source temperature.

---

[5] After all, there is no way from our observation point of proving that the distant galaxies are not moving away from us and that we simply occupy a privileged position.

[6] In the SR description, the recession velocity is constant, thus the CMBR sources were receding at the same velocity at the time the photons were emitted when the Universe was much smaller.

[7] We take the view that energy conservation implies that no closed loop process can result in the nett loss or gain of energy.

It is an error to add together the energy from all the observers in a scalar fashion and claim energy is lost. The observers are not in the same inertial frame and the straightforward addition of energy is not permissible. Yes, there is an energy difference between frames but it is not an energy loss. If all observers were instantaneously boosted by a velocity magnitude -$v$ in the appropriate direction, it is immediately clear there was no energy lost associated with the photons that were transferred whilst the velocity difference existed.

Cosmological events will map onto flat spacetime to exactly match the setup we have described and will therefore conserve energy. The energy difference is real but is completely accounted for by the transformation between inertial frames. There is no energy loss. In the SR description of the expansion described by (1), the recession velocity is constant with time. The idea that photons are losing energy over time as Hubble thought [6] is a misconception – yes, next year, the CMBR photon energy will on average be lower than this year, but they are photons originating from different inertial frames and cannot be directly compared. This is very different to GR where energy is really not conserved by the cosmological redshift because the same photons are steadily losing energy as they propagate [8, 9]. This is certainly a problem as the entire GR cosmology model is based on Energy conservation (the expanding Universe working against gravitation) and Friedmann's equation does not include this energy loss as a term.

This is not to say that GR cosmology is incorrect; far from it. It is merely being claimed that although GR specifies the mechanism and the model, SR describes the observations. The GR model and SR observation can therefore be made quite consistent by a decoupling process whereby the close link between GR and SR is broken. A coordinate mapping scheme of arbitrary complexity can be proposed where the energy differences in the differing redshift mechanisms of SR and GR are compensated for by apparent acceleration of cosmological bodies and the emergence of virtual forces [3]. In fact, the observed recession velocity of cosmological bodies is largely constant and there is little evidence of significant virtual accelerations. The puzzle is then how the GR model maps so simply onto the SR inertial observation frame. A further problem is that it is unclear if SR and GR can actually be decoupled when they are so closely intertwined through the geometry of spacetime. Nonetheless, any GR cosmological model must map onto the observation frame in a manner consistent with SR.

**THE EXPANSION ARISING FROM OBSERVATIONAL CONSIDERATIONS**

Although it is natural to try to explain the expansion of the universe through models, it should be born in mind that the apparent expansion can also emerge from SR by a requirement for observational consistency and GR is not actually necessary. If, as is believed, the Universe had a beginning in time, the Universe is compelled to expand in size at a rate equal to the speed of light if we are to avoid observational problems in SR. A qualitative justification for this conclusion is apparent when the look-back distance is considered. As time progresses, the look-back distance increases at the rate $c$. If the Universe did not expand, eventually the look-back distance would reach back the beginning of the Universe. A moment later, the look-back time exceeds the size of the Universe with a causal link to a point that does not exist; an obvious contradiction that results in an observational discontinuity at the origin of the Universe. This can be avoided if the Universe is expanding at the speed of light. The boundary of the Universe is thus moving away at the same rate as the look-back distance is trying to catch up. This is reflected in the redshift equation (1). After a time $\Delta T$ in the absorber frame, a source moving away at $v_H$ will later move at a velocity :

$$v_H' = \frac{r_d + v_H \Delta T}{T_e + \Delta T} = \frac{r_d}{T_e} = v_H . \qquad (2)$$

In the limiting case $r_d = c\, T_e$, the edge of the Universe is moving away at the speed of light and the same instant at $T = 0$ is in causal contact with the observer (all observers) at all times. All future space and time is therefore contained in the point of creation. This neatly explains why the Universe appears to be expanding at the rate $c$.

So what is the relevance of this conjecture to GR? GR is certainly the correct description of gravitation, but if the local matter inhomogeneity is removed by smearing particles evenly all over space then, we can argue, by symmetry, the net gravitational influence at any point should be zero. In a sense we are applying the cosmological principle, and if we try to model this we are immediately led to the standard GR models – there is no alternative (applying the cosmological principle to the SR equations gives rise to Milne cosmology and implies an infinite Universe). However, we are not concerned with models. The stress-energy tensor will not be zero, but the curvature will be very close to zero because of the extremely low matter density. In the weak-field limit the gravitational influence can be estimated using Poisson's equation which shows that spacetime topology can correctly be approximated to flat. If one were to look at the a point where the net force was zero a moment later, after which the expansion has made the Universe a bit larger, the force is still zero. It is therefore not obvious in what sense work has been done by the expansion. One must therefore be very careful when claiming the expansion does work against gravity. Thus in the SR formulation of the expansion, we do not even have to be concerned with the work done by the expansion.

**CONCLUSION**

We have described a model-free description of the expansion based solely on consistency. This is in many ways a complementary description to GR cosmology and in no way invalidates the established explanation – we are merely claiming GR cosmology is not the only explanation for observational data. SR already deals nicely with concepts like the stretching of supernovae light curves and apparent superluminal motion, but there are certain problems in supernovae data that need to be investigated, but it must be remembered that the SR description can be adapted in various ways to deal with this. Of course, by reverting to SR, we have in some ways taken a retrograde step as the Universe becomes even more mysterious to us. This is a heavy price to pay for the restoration of the principle of energy conservation. At least with the GR model, there was something that could be visualised: SR is a descriptive framework of rules, not a model. SR explains how the Universe is, but not why it is. We have no idea even about the basics - why must the speed of light be constant in each inertial frame in the absence of gravitation? Now we are claiming as an additional rule that the Universe will not permit space-time discontinuities. Why not? Any answer would merely be philosophical speculation, but it is hard to get away from the notion that the Universe is a unified structure in space and time and our problems in comprehension arise because the photon exchange mechanism grants us limited access to the Whole, much less that is necessary to understand the dynamics of the Universe. Our theories and equations are then simply an expression of our limited vision[8].

**REFERENCES**

[1] D. N. Spergel, L. Verde, H. V. Peiris, E. Komatsu, M. R. Nolta, C. L. Bennett, M. Halpern, G. Hinshaw, N. Jarosik, A. Kogut, M. Limon, S. S. Meyer, L. Page, G. S. Tucker, J. L. Weiland, E. Wollack and E. L. Wright, "First-Year Wilkinson Microwave Anisotropy Probe (WMAP) Observations: Determination of Cosmological Parameters", Ap.J.S., **148**(1): 175-194 (2003)
[2] L. Covi, "Status of Observational Cosmology and Inflation", *Physics in Collision*, Proceedings of the XXIII International Conference. Edited by S. Riemann and W. Lohmann. SLAC eConf http://www.slac.stanford.edu/econf/C030626, p.67 (2003)
[3] A. Macleod, "An Interpretation of Milne Cosmology", arXiv:physics/0510170 (2005)
[4] T. M. Davis and C. H. Lineweaver, "Expanding Confusion: Common Misconceptions of Cosmological Horizons and the Superluminal Expansion of the Universe", Publications of the Astronomical Society of Australia **21**(1): 97 – 109 (2004)
[5] M. J. Chodorowski, "Cosmology under Milne's Shadow", arXiv:astro-ph/0503690 (2005)
[6] E. Hubble, *The Realm of the Nebulae*, Yale University Press (1936)
[7] A. Macleod, "Redshift and Energy Conservation", Section 2, arXiv:physics/0407077 (2004)
[8] E. R. Harrison, *Cosmology, The Science of the Universe*, Cambridge University Press (1981)
[9] P. J. E. Peebles, *Principles of Physical Cosmology*, Princeton University Press (1993)
[10] H. Spinrad, "Galaxies at the limit: The epoch of galaxy formation", in *The Farthest Things in the Universe*, Eds J. M. Pasachoff *et al*, Cambridge University Press (1994)
[11] H. Spinrad *et al*, Publ. Astron. Soc. Pac. **97**, 932 (1985) Electronic Version: http://cdsweb.u-strasbg.fr/viz-bin/Cat?J/PASP/97/932

---

[8] If that is true, an empirical-inductive approach to theory–building is doomed to failure and we can never create a comprehensible model for the Universe. There is however no reason why the hypothetico-deductive method cannot be applied to produce an abstract mathematical description that completely matches observation.

**APPENDIX 1**

The critical data for any cosmological model is around the region from $z \sim 1$ to $z \sim 2$. There is still a paucity of supernovae Ia data from this epoch, but many quasars and radio galaxies have been observed. Though by no means standard candles because of a wide scatter of intrinsic luminosities and possibly evolution effects, we might still get a 'preview' of what we might expect from emergent supernovae data by looking at this data (following an earlier analysis by Spinrad [10]).

The SR luminosity distance, $D_L$, is [5]:

$$D_L = (1+z)^2 \frac{vT}{1+\beta} \quad \left[\equiv cT(z + z^2/2)\right], \quad (1.1)$$

where $T$ is the current age of the Universe, $v$ is the recession velocity, $\beta$ is $v/c$ and $z$ is the redshift. All the sources in the 1985 update of the Revised Third Cambridge Catalogue (3CR) [11], except for the single BLac, were included in the analysis. The catalogue entries are categorised into quasars, galaxies and N galaxies. An absolute visual magnitude of –22 was arbitrarily applied to the galaxies and –26 for the quasars. The results are shown in Fig. 1.1.

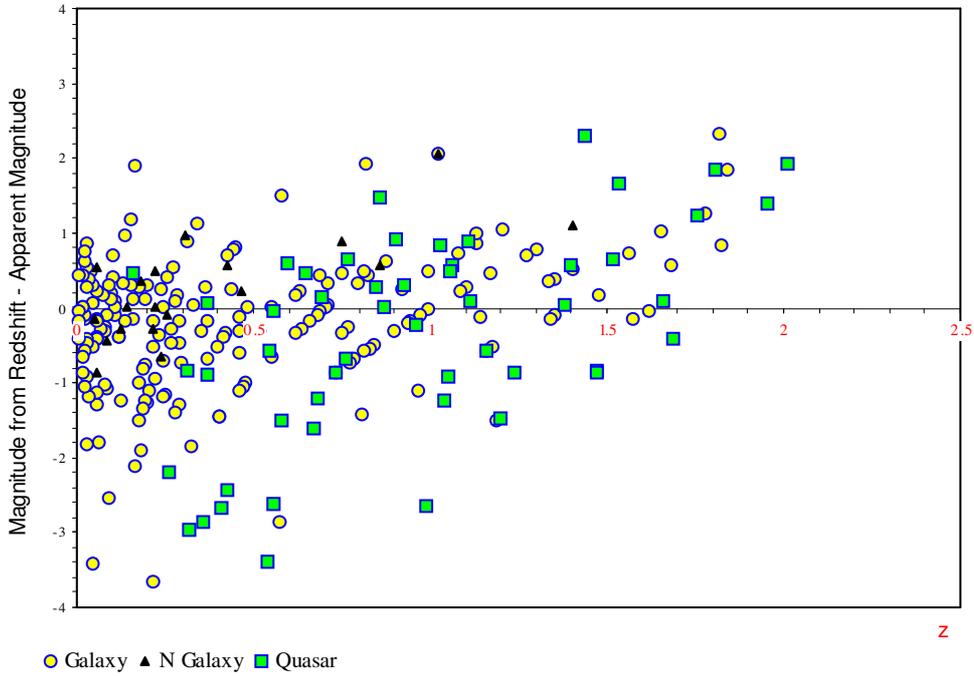

**Figure 1.1:** Modelling radio galaxies and quasars from the 3CR catalogue with the SR apparent distance – redshift formula. The deviation assuming all galaxies have an intrinsic magnitude of –22 and the quasars have a value of –26 is plotted. An age of 13.7 Gyr for the Universe is assumed.

Superficially, there is some cause for concern with reference to SR predictions: the distant sources are brighter than expected. However, it should be noted that there is a very wide scatter over the entire range that is not observed at high redshift. It is therefore possible that the effect is the result of a selection bias that picks out only the brightest sources. In any case, a simple SR approach to cosmological observations around $z \sim 1$ is probably inadequate.

# APPENDIX 2

We expect consistency and conservation when we map cosmological observations onto a flat Universe and we can prove this is indeed the case. The basic problem can be represented by a rod expanding with the Universe (Fig. 2.1).

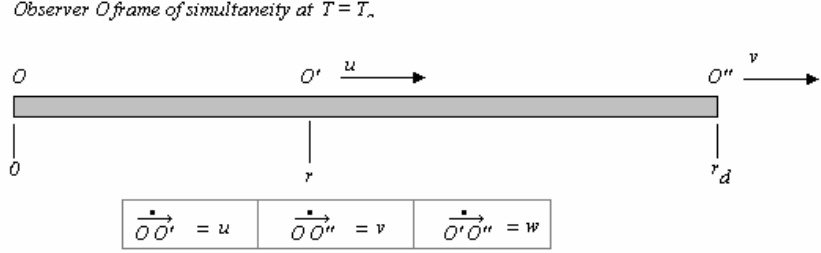

**Figure 2.1:** A rod of length $r_d$ in the frame of simultaneity of the observer at $O$, the other end of the rod. $u$ and $v$ are the apparent velocities with respect to $O$ and $w$ is the velocity of the end of the rod as seen by $O'$.

An event at the end of the far end of the rod, $(r_d, T_e)$ is perceived at positions 0 and $r$ as events $(0, T_e + r_d/c)$ and $(r, T_e + (r_d - r)/c)$ respectively (again coordinates with respect to $O$). Since the recession velocity is defined as the proper distance at emission divided by the time of emission (both in the observer frame), it immediately follows that

$$v = \frac{r_d}{T_e}, \qquad (2.1)$$

and

$$u = \frac{r}{T_e + (r_d - r)/c}. \qquad (2.2)$$

Note these apparent recession velocities do not change over time. By assuming all sources started at the same point and that there is a difference in the rate at which time passes between moving frames, it can be argued that the time in the $r$ frame when the passing photon is received is $\dfrac{T_e + (r_d - r)/c}{\gamma_u}$, and this can be verified by applying a velocity boost of $-u$. The coordinates of the $O$ event $(r_d, T_e)$ and the connected events at $r$, and 0 are in the $O'$ frame (in the order from L→R):

$$\left(-\gamma_u u(T_e + r_d/c), \gamma_u(T_e + r_d/c)\right), \qquad (2.3a)$$

$$\left(\gamma_u(r - uT_e - u(r_d - r)/c), \gamma_u(T_e + (r_d - r)/c - ur/c^2)\right), \qquad (2.3b)$$

and

$$\left(\gamma_u(r_d - uT_e), \gamma_u(T_e - ur_d/c^2)\right). \qquad (2.3c)$$

It is straightforward to verify that $\gamma_u(T_e + (r_d - r)/c - ur/c^2) = \dfrac{T_e + (r_d - r)/c}{\gamma_u}$. The distance travelled by the photon is the intermediate frame is $\gamma_u(r_d - r)(1 + u/c)$. The apparent expansion velocity of the end of the rod as seen form the intermediate position, $w$, is derived by recognising that the photon travel distance is the arrival time times $w$ divided by $(1+w/c)$ (all in the intermediate observation frame):

$$\frac{w}{1+w/c} = \frac{\gamma_u(r_d - r)(1 + u/c)}{\gamma_u(T_e + (r_d - r)/c - ur/c^2)} = \frac{(r_d - r)(1 + u/c)}{(T_e + r_d/c - r/c(1 + u/c))}. \qquad (2.4)$$

Simplifying:

$$w = \frac{(r_d - r)(1 + u/c)}{T_e - u r_d / c^2}.\tag{2.5}$$

We can manipulate this expression. First replacing $r_d$ with $vT_e$:

$$w = \frac{(v - r/T_e)(1 + u/c)}{1 - uv/c^2}.\tag{2.6}$$

Expanding

$$w = \frac{v - u\left(\dfrac{r}{uT_e} - \dfrac{v}{c} + \dfrac{r}{cT_e}\right)}{1 - uv/c^2} = \frac{v - u\left(1 + \dfrac{r_d - r}{cT_e} - \dfrac{v}{c} + \dfrac{r}{cT_e}\right)}{1 - uv/c^2}$$
$$= \frac{v - u}{1 - uv/c^2}.\tag{2.7}$$

Thus the apparent velocities are completely compatible with the relativistic sum of velocities, i.e. $v$ is equivalent to $w$ when viewed form a frame moving at $u$. This is true $\forall\, r$ such that $0 \leq r \leq r_d$.

Energy conservation must hold. This is conveniently expressed as

$$1 + z_v = (1 + z_u)(1 + z_w).\tag{2.8}$$

Because $(1 + z_\Omega)^2$ is defined as $(c+\Omega)/(c-\Omega)$, the expression is

$$v = \frac{u + w}{1 + uw/c^2},\tag{2.9}$$

which of course is completely equivalent to the 'sum of velocities' expression above.

Consistency is ensured. The definition of recession velocity given in equation (1) is completely equivalent to proper velocities in SR.